# HBT-Analyzer – Particle Correlations Analysis Toolkit

Piotr Krzysztof Skowroński for ALICE Collaboration
*CERN, Genève 23, CH-1211, Switzerland and*
*Warsaw University of Technology, Faculty of Physics, ul. Koszykowa 75, 00-662 Warsaw, Poland*

The HBT-Analyzer is an universal tool for particle correlations analysis under the ROOT environment. It provides an efficient mixing mechanism, a wide range of correlation and monitoring functions, and a set of cuts that are applicable on different levels of analysis. Thanks to an object oriented design it is very easily extensible for particular user needs. It also enables easy resolution studies of Monte-Carlo simulated data. The general concept and design is presented as well as example analysis results obtained within ALICE experiment.

## 1. INTRODUCTION

This paper describes a package for particle correlations analysis developed within the ALICE [1] offline framework [2]. Correlations are used in particle physics to measure the size of emission volume [3]. It has deep analogy with Hanbury-Brown and Twiss (HBT) method used in astronomy for star size measurement [4]. This name is also commonly used in High-Energy Physics. HBT-Analyzer is applicable in any-multi event analysis where all particle combinations are considered, for example invariant-mass analysis.

## 2. GOAL OF HBT ANALYSIS

The output of HBT analysis is a Correlation Function (CF) of some variable $Q$ (e.g. $Q_{inv}$, $Q_{Out}$, $Q_{Side}$, $Q_{Long}$ or $M_{inv}$). CF is a histogram created from the division of two other histograms, further called numerator and denominator. Numerator is obtained from $Q$ values calculated for pairs of particles (or triplets) coming from the same events. Denominator is calculated the same way, however particles are taken from different events. This process is called mixing.

## 3. ALGORITHM

Algorithm is rather simple and consists of the following steps:
1. Loop over events (I)
2. Loop over particles from event I
3. Loop over events (II)
4. Loop over particles from event II
5. Checking cuts
6. Calculation of desired $Q$ value
7. If particles come from the same event numerator is filled, or denominator in the other case.

The goal was implementation of this algorithm in such a universal way that it can be used in any kind of analysis.

During the design process the following use-cases were taken into account:
- calculation of many correlation functions in single analysis pass
- calculation of the same kind of function but with different cuts in single analysis pass
- in Monte-Carlo data analysis comparison between original events and events after detector simulation and reconstruction
- monitoring of single particle spectra
- resolution plots.

Performance was a very important objective during software design and implementation. This is especially important in such analysis since it is remarkably time consuming (quadruple loop), and each unnecessary calculation or call inside the inner loop transforms into hours or even days of computing time.

It is designed so that the user can steer the analysis process by applying cuts at almost any level of analysis, starting at data reading, through the mixing procedure and finally in each calculated function.

## 4. ARCHITECURE

Architecture and object hierarchy is shown in Fig. 1. The central object of the package is **analysis** (class *AliHBTAnalysis*). It takes data from a **reader** (derivative of *AliHBTReader* class). A user can plug into the analysis arbitrary number of **correlation** and **monitoring functions**. **Particle cut** (*AliHBTParticleCut* class) and **pair cut** (*AliHBTPairCut* class) can be applied on different levels of analysis. **Pair cut** is a data member of **analysis** and **correlation function** classes. **Pair cut, reader** and **monitor functions** use **particle cut**. Data are stored in and exchanged via internal structures: **particle** (class *AliHBTParticle*), **event** (class *AliHBTEvent*) and **run** (class *AliHBTRun*).

### 4.1. Functions

The function is an object that calculates one histogram. In experimental data analysis the most important function is the one that fills the numerator and denominator histograms for pairs of reconstructed particles. However, there are more kinds of functions needed:
- resolution functions (especially required in Monte Carlo data analysis)
- monitoring functions – enable to see distributions of particles properties used in analysis (influence of cuts)

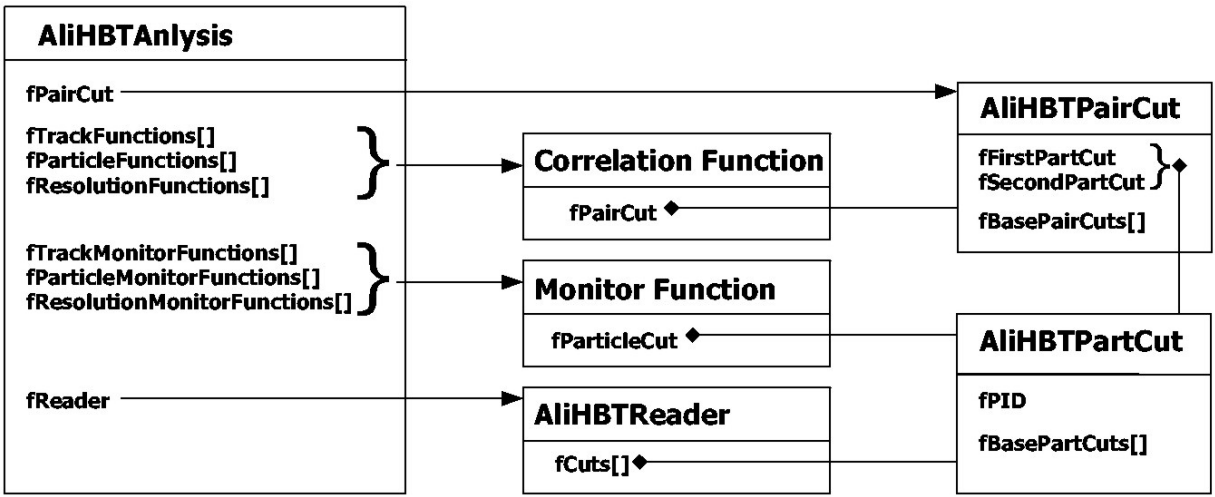

Figure 1: Architecture: class dependence schema

All these functions have different input parameters and need to be handled separately in analysis. Functions in the HBT-Analyzer can be systematized according to two criteria: the functionality that they serve and the dimensionality. Using the functionality criterion, correlation and monitoring functions are distinguished. Accordingly to the dimensionality functions are categorized into 1D, 2D and 3D.

### 4.1.1. Correlation functions

Currently two kinds of correlation functions are implemented. They can be one- or two-pair, depending on the input they require to fill one entry. One-pair function is the most important in experimental data analysis and is a base class for such one as CF($Q_{inv}$). Two-pair function is designed for resolution analysis, where the comparison of value calculated for pair of simulated particles with one made of reconstructed tracks is needed. They are also used in so-called weight algorithms [3]. Implementation of one- and two-triplet functions is foreseen in the future, and it will enable handling three particle correlations.

The fact that all of functions mentioned above can be one-, two- or three-dimensional, imposes class hierarchy as shown in Fig. 2. *AliHBTAnalysis* class depends only on interfaces defined by *AliHBTOnePairFctn* and *AliHBTTwoPairFctn* classes. They define pure virtual methods
- *ProcessSameEventParticles*
- *ProcessDiffEvenOarticles*
- *Init*
- *WriteFunction*

The first two of these methods differ in number of input parameters. *AliHBTOnePairFctn* requires one-pair whereas *AliHBTTwoPairFctn* needs two pairs.

All functionality that is dependant on function dimension is implemented in *AliHBTFunction1D*, *AliHBTFunction2D* and *AliHBTFunction3D* classes, which in turn inherit from *AliHBTFunction* class. *AliHBTFunction* class implements methods common for all functions and fixes user interface for (re)naming, writing, resetting, scaling, etc. routines.

The classes that constitute the bottom layer of the

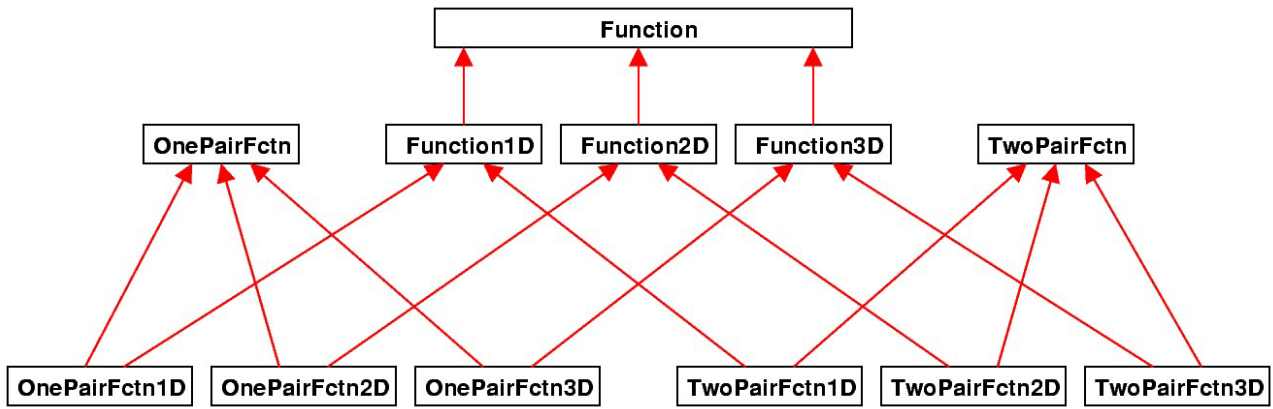

Figure 2: Correlation functions inheritance schema. In all classes AliHBT prefix is skipped.

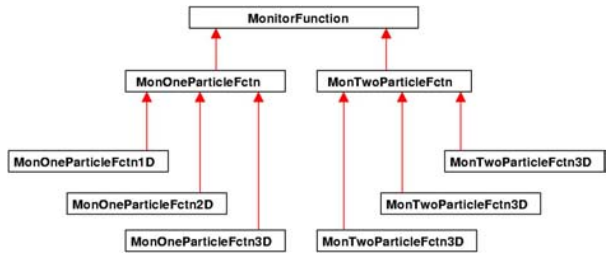

Figure 3: Monitoring functions class herarchy,

diagram are base ones for concretized user classes. Creating a new correlation function requires implementing only two methods in most cases:
- *double GetValue(AliHBTPair*)*
- *TH1* GetResult()*

The first one calculates the correlated value for a pair and the second one returns a result histogram. HBT-Analyzer already implements a wide set of most common correlation functions like $Q_{inv}$, $Q_{Out}$, $Q_{Side}$ or $Q_{Long}$.

### 4.1.2. Monitoring functions

Monitoring functions are designed to create distributions of particle and pair properties that are used in analysis. This functionality is especially desired when the user requires knowing the influence of applied cuts on a given distribution. Currently there are one- and two-particle monitoring functions implemented in the HBT-Analyzer.

The inheritance tree of monitoring functions (Fig.3) is simpler than the hierarchy of correlation functions. There are no separate base classes for each dimension, since they would be irrelevant in this case. They would be completely empty with the definition of just one histogram each.

All classes in Fig.3 are abstract and user functions must be derived from the leaf classes.

### 4.2. Analysis object

User can set six types of functions i.e. three types for correlation and monitoring each, namely for:
- reconstructed tracks,
- simulated particles,
- reconstructed tracks and simulated particles.

The last item corresponds mainly to resolution analysis but is also used in weight algorithms. Therefore, a more general term is used. In the real data analysis only reconstructed track functions are calculated. Two other types are used only in Monte–Carlo data analysis.

The user triggers analysis by calling the *Process* method with an option indicating which data are to be used in the analysis: simulated, reconstructed or both. In the case of both being used simultaneously, the *reader* must ensure that $n^{th}$ particle corresponds to $n^{th}$ reconstructed track in a given event.

In a general case, all particles are used in analysis and every particle is mixed with all the others. In order to speed up computation time, the user can set a pair-cut object in the analysis object that filters out pairs before passing them to functions. For optimization reasons this cut is checked progressively. The first check is performed immediately when the first particle is picked up. If the particle does not pass the check then it does not make sense to loop over all the other particles. The next check is performed when the second particle is picked from an event, and at last the pair properties are verified. Pair cut is designed in a way that allows checking the pair properties only without examining each particle in a pair to avoid repeating of the tests.

In the case of non-identical particle analysis, or more generally, in the case of mutually exclusive cuts on individual particles (a particle cannot be accepted by both cuts) faster algorithm is used. In such case pre-selection of particles is performed. In the outermost event loop all particles are checked if they pass any of the individual cuts in a pair. Pointers to particles that passed the first or second particle cut are stored in two separate arrays. Such pre-selected particles are mixed and numerators are filled. Arrays of particles that passed a cut on second particle are buffered in a FIFO queue. Particles from the buffer are mixed with the array of first cut particles and denominators are filled. The size of the queue is adjustable by the user. The choice of mixing algorithm is made automatically by an analysis object on the basis of the particle-cut properties.

One would think that the pre-selection technique is also faster in a general case and that it is enough to make sure that a particle is not mixed with itself. However, in this case the optimal way is particle selection on the level of reading (see section 4.3).

### 4.3. Readers

Reader is an object that provides data to analysis. HBT-Analyzer is easily customized to any input data format by creating a specialized reader that inherits from class *AliHBTReader*. This base class defines pure virtual interface that allows implementation of buffering as well as non-buffering readers, depending on particular user needs.

The base reader class implements functionality for particle filtering at reading level. The user can set any number of particle cuts in a reader, and the particle is read only if it fulfills the criteria defined by any of them.

It also has a feature that allows specifying multiple data sources, which can be sequentially read. Many analysis objects can use one reader instance, if needed.

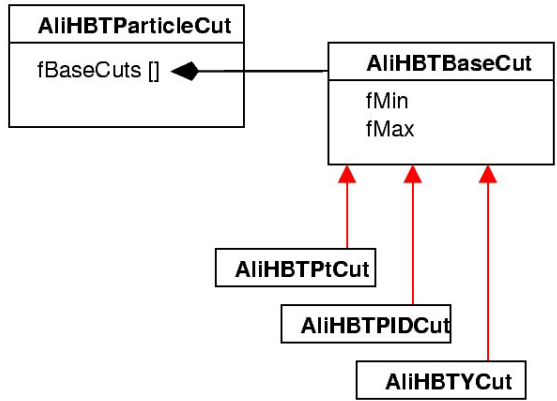

Figure 4: Particle cut – classes diagram

### 4.4. Cuts

Currently the HBT-Analyzer implements two kinds of cuts: particle and pair cut. Cut object defines the ranges of many properties that a particle or a pair may posses and it also defines a method, which performs the necessary check. However, usually a user wants to limit ranges only of a few properties. For speed and robustness reasons, the design presented in Fig. 4 was developed.

The particle cut object has an array of pointers to particle base cuts (class *AliHBTBaseCut*). The number of entries in the array depends on the number of the properties set by the user. The base cut implements checks on a single property. It implements maximum and minimum values and a virtual method *Pass* that performs a range check of the value returned by pure virtual method *GetValue*. Implementation of a concrete base cut is very easy in most cases: it is enough to implement *GetValue* method. The HBT-Analyzer already implements a wide range of base particle cuts, and *AliHBTBaseCut* class has a comfortable interface for setting all of them. For example it is enough to invoke the *SetPtRange(min,max)* method and behind the scenes a proper base cut is created and configured.

All objects, which have a member pointer to a particle cut, by default create an empty cut (class *AliHBTEmptyParticleCut*). This is a derived class of *AliHBTParticleCut* that accepts any particle.

The pair cut follows the design of the particle cut. In addition to an array of pointers to the base pair cuts it has two pointers to particle cut, one for each particle in the pair. An empty pair cut is implemented as well. In addition, all objects that have a pair cut data member by default create the empty cut.

### 4.5. Pair

The pair object points to two particles and implements a set of methods for the calculation of pair properties. The pair buffers calculated values and intermediate results for performance reasons. This applies to the quantities whose computation is time consuming and also to quantities with a high reuse probability. A Boolean flag is used to mark the already calculated variables. To ensure that this mechanism works correctly, the pair internally sets and reads the values of its variables calling its own methods, instead of accessing the variables directly.

Pair object points to another pair with swapped particles. The existence of this feature is connected with the mixing implementation: when particle A is mixed with B, swapped pair is not going to come. In non-identical particle analysis their order is important, and pair cut may reject a pair while a reversed one would be accepted. In order to solve this problem, pair cut checks swapped pair as well, if a regular one is rejected. This solution automatically takes advantage of the buffering technique.

## 5. SUMMARY AND FUTURE DEVELOPMENT PLANS

HBT-Analyzer is a universal and robust tool for particle correlation analysis. It can be used by any experiment providing the proper reader object is implemented. The design ensures that it is very easy customizable and extendible to new functions and cuts.

Special care is taken for the performance of the package. Several techniques like buffering, specialized mixing, and cuts with flexible lists of properties are used.

ALICE successfully uses this tool in studies of detector and reconstruction software capabilities [6]. Design ensures that we are already prepared for real data analysis. Example plot is presented in Fig.5, where is $Q_{inv}$ correlation function for events generated with the Monte-Carlo generator (dots) and this function for tracks reconstructed with Alice Time Projection Chamber (squares) and Inner Tracking System (triangles).

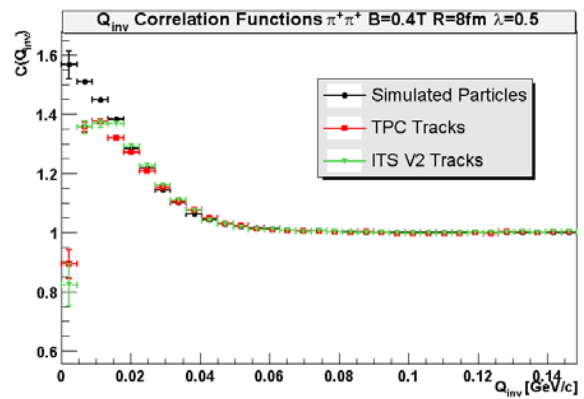

Figure 5: Example results: $Q_{inv}$ correlation function for $\pi^+\pi^+$

The HBT-Analyzer is still being developed and extended. The main features that are planned for implementation in the near future are event cut, three particle and azimuthally sensitive analysis. Software and documentation are available on [7].